# Radiation-technological processes under intense irradiation of a complex fractal medium

**Zatsepin Anatoly Fedorovich[1], Oksengendler Boris Leonidovich[2],**
**Suleymanov Sultan Khamidovich[3], Ibrokhimov Rakhmatullo[4], Nigora Turaeva[4]**
*[1]Professor, Candidate of Technical Sciences, Leading Researcher.*
*Physico-Technological Institute, Ural Federal University,*
*Yekaterinburg, 620078 Russia*
*[2]Professor, Doctor of Physical and Mathematical Sciences, Leading Researcher.*
*Institute of Materials Science, Academy of Sciences of the Republic of Uzbekistan,*
*Tashkent, 102226 Uzbekistan*
*[3]Candidate of Physical and Mathematical Sciences, Head of Laboratory,*
*Institute of Materials Science, Academy of Sciences of the Republic of Uzbekistan,*
*Tashkent, 102226 Uzbekistan*
*[4]Junior Researcher,*
*Institute of Materials Science, Academy of Sciences of the Republic of Uzbekistan,*
*Tashkent, 102226 Uzbekistan*

*e-mail: oksengendlerbl@yandex.ru

**Abstract.** A fracton model of a complex medium has been constructed to study physicochemical processes for radiation-technological purposes that occur under intense irradiation. The possibility of shock-wave generation, as well as of percolation phenomena involving fractal aggregates, has been studied in the search for an ideal thermal insulator.

## 1. Introduction.

The “Complexity” concept, which emerged in the 1970s, exerted a profound influence on materials science, so strong that it even became t he basis for answering the question: “How does Nature work?” [1] . At the beginning of the 21st century, the ideas of “Complexity” expanded into the field of radiation physics, where their appearance proved particularly revolutionary, due to their strong influence on the openness of systems (both living and non-living Nature), always characterized by a pronounced manifestation of nonlinearity and nonequilibrium [2–3]. In addition to these two properties, “Complexity” must also account for a fundamental revolution in materials science itself: the emergence of numerous material objects whose atomic structural features turned out to be more important than the features of their electronic structure. The development of these representations revealed the optimal structure of the entire methodological scheme of "Complexity": its reliance on two concepts (synergetics and synergistics), which formed five paradigms (self-organization, dynamic chaos, self-organized criticality, network connections, non-additivity) and materialized in concrete physico-mathematical models (Fig. 1). Using both analytical and computational methods, this approach made it possible to resolve numerous paradoxes discovered experimentally across virtually all natural sciences — in physics, chemistry, biology, and medicine . Applied to radiation physics, the "Complexity" methodology enabled the elucidation of highly complex radiation effects that are currently relevant and associated with exposure to electromagnetic and ultrasonic radiation, as well as to corpuscular irradiation, including high-intensity fluxes. Particularly interesting are the effects under combined action of radiation factors (high-energy ions, lasers, synchrotrons, and Large Solar Furnaces). With the introduction of the "Complexity" methodology, it became possible to achieve predictable radiation technology in many difficult cases.

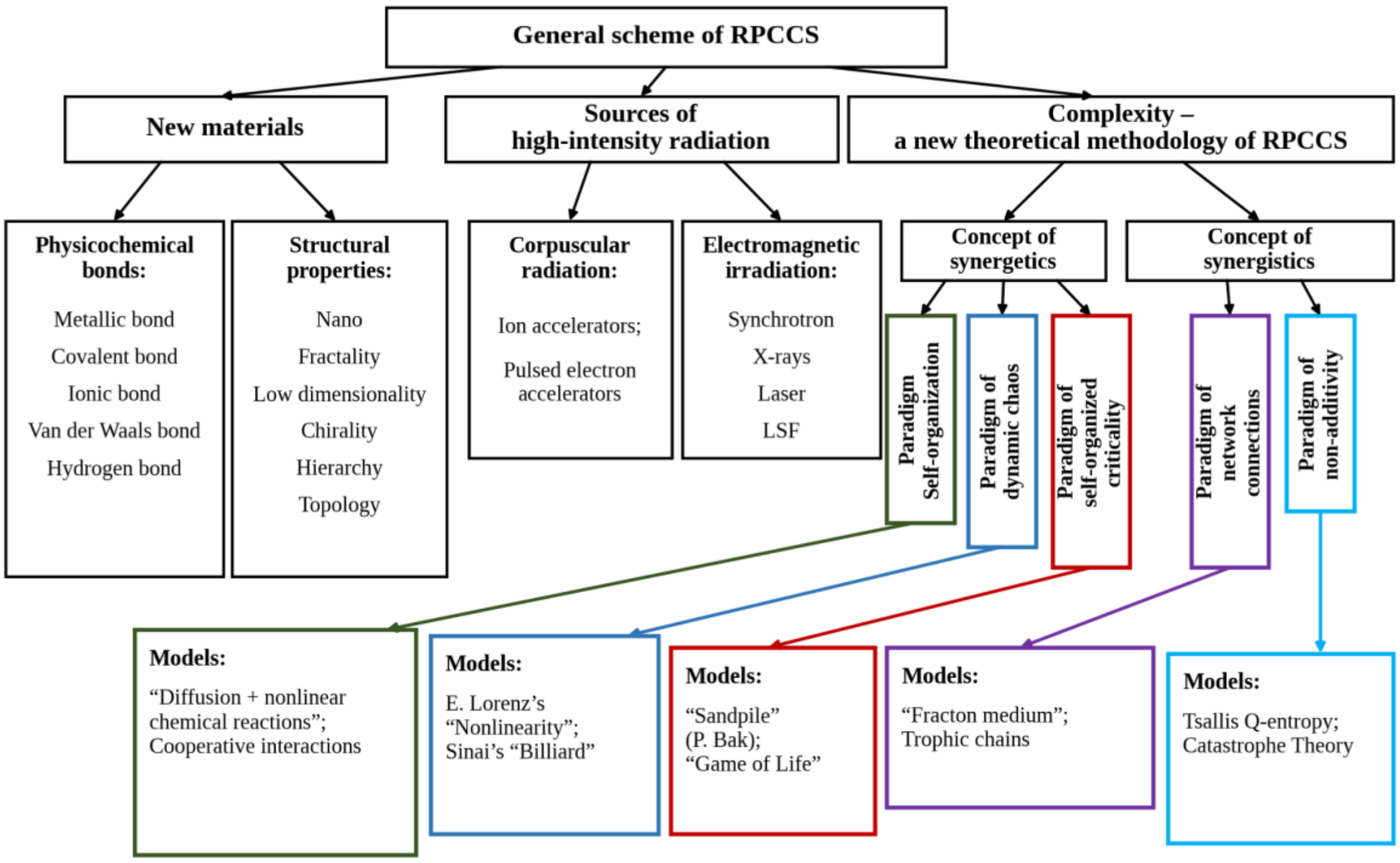

**Fig. 1** Extended scheme of the radiation physics of condensed complex systems (RPCCS)

Analyzing the general scheme of modern radiation physics of condensed complex systems (RPCCS), one can see that the complexity of the radiation response of the systems under study arises from three aspects: first, the structure of complex systems (left part of the figure); second, the combined nature of the irradiation factors, which become especially pronounced at high radiation intensities (central part of Fig. 1); and third, the possibility of describing the entire set of manifested radiation properties within the framework of the “Complexity” paradigm and all its hierarchical subdivisions (right part of Fig. 1).

Focusing on the capabilities of the radiation technology of the Large Solar Furnace (LSF), one can base the analysis on a characteristic irradiation scenario—namely, exposure to a so-called fracton medium.

## 2. Features of the radiation response of network molecular structures under radiation technology using intense radiation sources.

It is important to emphasize that the models themselves concisely embody both the characteristics of the target materials and the properties of specific types of radiation exposure. In principle, radiation transfers its energy to matter through five channels: elastic scattering, ionization, heat, shock waves, and elastic waves (phonons). The specifics of radiation exposure in the Large Solar Furnace (LSF) lie in the fact that its irradiation factors, when combined, give rise to five influencing factors: ionization, excitation of chemical bonds (phonons), heat, and temperature gradients. As for shock waves (SW), which are absent from the conventional list of radiation exposure factors in the LSF [2–3], it is not immediately clear what mechanisms could generate SW in this case. Nevertheless, this is a very important channel, since irradiated materials are almost always metastable, and shock waves can modify or even eliminate this metastability (the so-called “radiation shaking” [4]).

In this regard, the aim of the present article is to identify mechanisms for the generation of shock waves and to explore certain aspects of their application in radiation technologies involving intense irradiation (including the LSF).

### 2.1. Features of the fracton medium in the study of radiation effects.

In recent years, objects that can be classified as network systems have attracted increasing interest. These systems model an extremely wide range of situations, including those of particular relevance to radiation materials science (Fig. 1). For radiation materials science in the context of the Large Solar Furnace (LSF), network systems are relevant as certain quasi-one-dimensional chains possessing fractal dimensionality and varying topological connectivity indices. Simplified, this can be interpreted as follows: we are dealing both with continuous quasi-one-dimensional chains of fractal dimensionality and with fractal aggregates [2–3]. In the latter case, the medium contains chain segments that exhibit a set of fractal dimensions; more simply, these are two- or three-dimensional “fractal clusters” connected by quasi-one-dimensional fractals or even terminated by dead ends. Studies show that these are entirely real systems characteristic of irradiated objects in the LSF, and in this context they display exceptionally interesting and distinctive properties. Given the particular significance of such structures in conventional materials science and their special promise under radiation technology conditions, properties not previously reported anywhere, we now consider these features in greater detail.

Thus, fractal aggregates possess a unique ability to accumulate energy from radiation in the LSF, while the translation of these **fracton** excitations along the entire object (the chain) is hindered. This is a fundamental property of the material as a whole, commonly referred to as a “fracton medium”. The characteristic appearance of a fracton medium is shown in Figs. 2 and 3 [5].

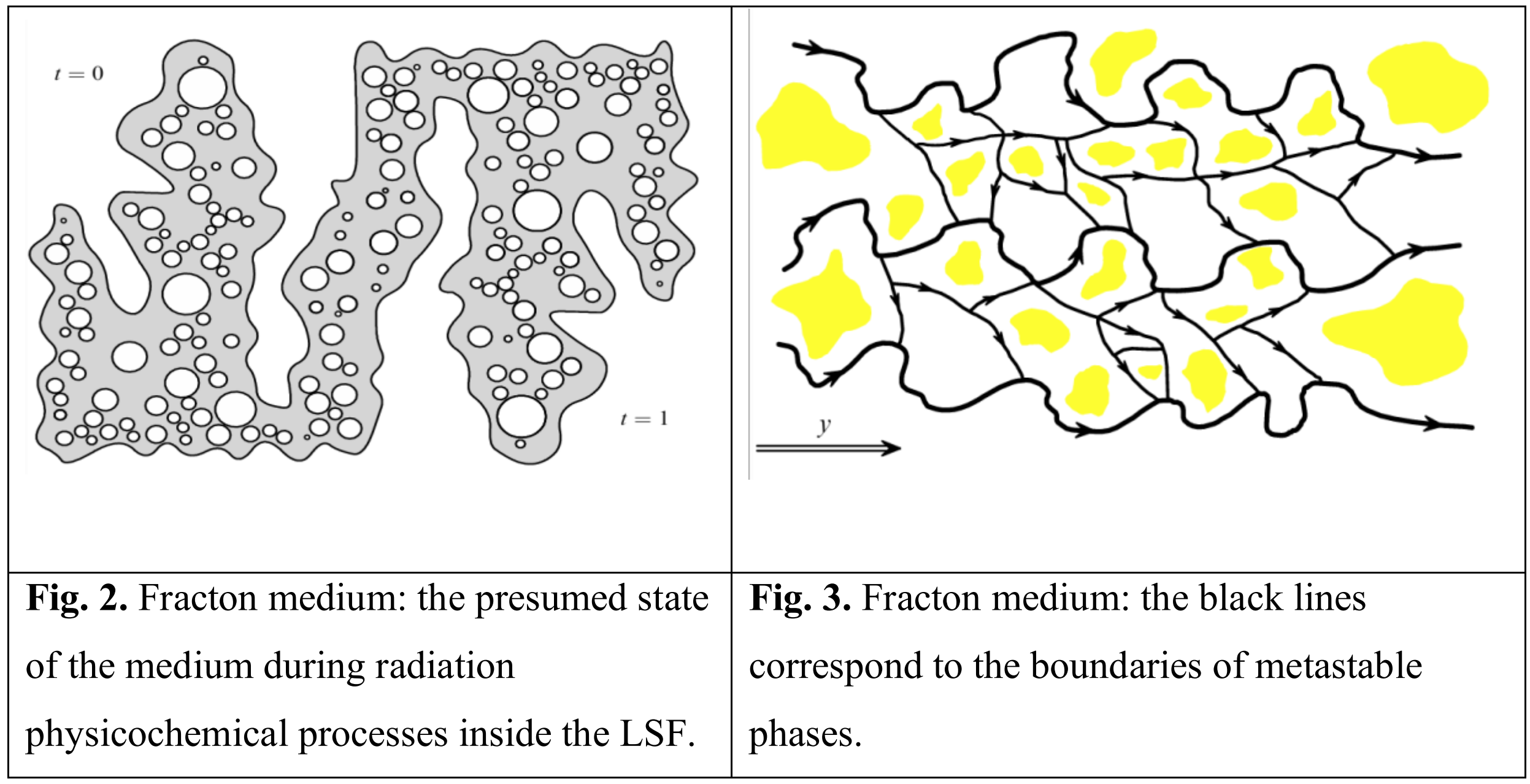

| **Fig. 2.** Fracton medium: the presumed state of the medium during radiation physicochemical processes inside the LSF. | **Fig. 3.** Fracton medium: the black lines correspond to the boundaries of metastable phases. |
|---|---|

The properties of non-integer connectivity, fractality of chain elements, metastability, and other characteristic features of complex systems are indicated in Fig. 1. Investigating electron–molecular relaxations in such systems requires specialized tools: integrodifferential equations with fractional exponents, nonlinear percolation theories, methods of fractal dynamics, and others . Naturally, in the case of multifactorial radiation exposure, taking into account the openness of the systems, their complete description is rather complex. However, if we restrict ourselves to the mechanisms of radiation action on fracton media, it is often possible to use more familiar mathematical language.

**2.2. Generation of shock waves in a fracton medium.** A fracton medium possesses a number of other specific properties; in particular, it is characterized by extremely low thermal conductivity, that is, it acts as a unique thermal insulator. Considering a fracton medium, one can introduce a fracton temperature for fractal aggregates, which can increase by absorbing radiation energy from the broad-spectrum LSF. Analysis shows that, in several relevant cases, the equation for the temperature growth of a fractal aggregate has the form of a time-dependent curve with a maximum, featuring a markedly elongated relaxation channel over time, for example:

$$T_f \sim \frac{1}{t^{1+\alpha}}, \quad (1)$$

where $\alpha < 1$ [5].

Thus, there is both an inflow and an outflow of energy, and if the energy inflow occurs sufficiently fast and collectively, the situation is resolved by a thermal explosion, in which an SW with a number of characteristics is generated.

In a rough approximation, this situation can be modeled as a "point explosion." The development of the SW front in time is then described by expressions for the SW front radius ($R$) and the pressure at the SW front ($\rho$). Accordingly, we have:

$$R \simeq (E_+/\rho_0)^{1/5} t^{2/5}; \quad (2)$$

$$p \simeq \rho_0 [E_+/(\rho_0 t^3)]^{2/5}. \quad (3)$$

In these formulas, $E_+$ is the energy reserve of the excited region at time $t = 0$, and $\rho_0$ is the density of the medium.

Next, it is not difficult to obtain an important relation linking the pressure at the front and the radius of the SW:

$$p \simeq (E_+/\rho_0)/R^3. \quad (4)$$

Formula (4) is extremely interesting from the standpoint of a metastable medium and the possibility of eliminating metastabilities, the elements of such a medium. Indeed, if the metastability energy is Q, and the volume increase associated with this metastability is $\Omega_0$, then the pressure exerted on this metastability that is able to eliminate it satisfies the condition $p^* \geq \frac{Q}{\Omega_0}$. Then, throughout the spherical region (around the SW center) where the condition $p \geq p^*$ holds, the metastability in question will be eliminated. Using formula (4), we immediately find the radius of the region "cleaned" of all metastabilities of the type $\{ Q, \Omega_0 \}$:

$$R^* = \left[\frac{E_+}{\rho_0}\frac{\Omega_0}{Q}\right]^{1/3} \quad (5)$$

Knowing the quantity $R^*$,, it is easy to obtain (using the concept of the random-neighbor probability $\omega(R)$ [4]) the probability of "annealing" (by the radiation shaking mechanism) of the metastabilities in question:

$$W^* = \int_0^{R^*} \omega(R) \cdot d^3R \quad (6)$$

**3. Percolation properties in a fracton medium under irradiation.** The peculiarities of a fracton medium are such that percolation phenomena in it can occur at two scales. First, inside fractal aggregates, which are characterized by chains of different dimensionality, both connected to each other and containing breaks. Therefore, kinetic phenomena inside fractal aggregates may correspond to Figs. 3–5, and such a structure is ideally described by the bond approach, with chains containing "dead" ends.

Second, when kinetic processes concern the sample as a **whole**, percolation corresponds to the involvement of all fractal aggregates in realizing the integrity of the process under study in the given object. The random network model can also be used here, but the simpler, well-studied site model on a square lattice (Fig. 5) is also suitable:

$$x = N_B/(N_B + N_W)$$

As is known, the extensive body of percolation theory available to date indicates that integrity, interpreted as the formation of an infinite cluster, is observed when a certain concentration (probability) barrier is overcome in all problems, both "bond" and "site." A characteristic graph of such a dependence (the percolation probability) is shown in Fig. 5.

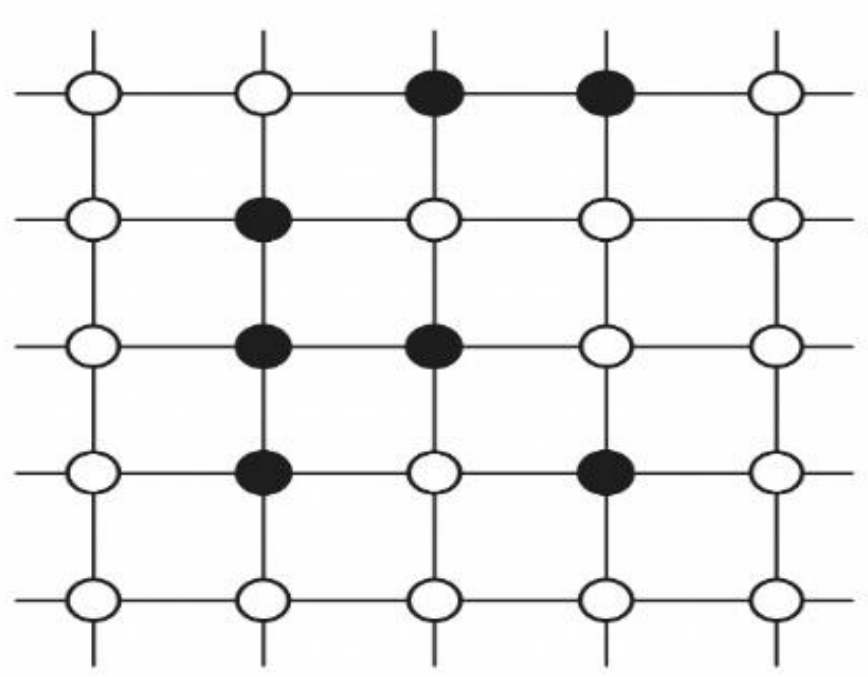

**Fig. 4.** Site problem on a square lattice.

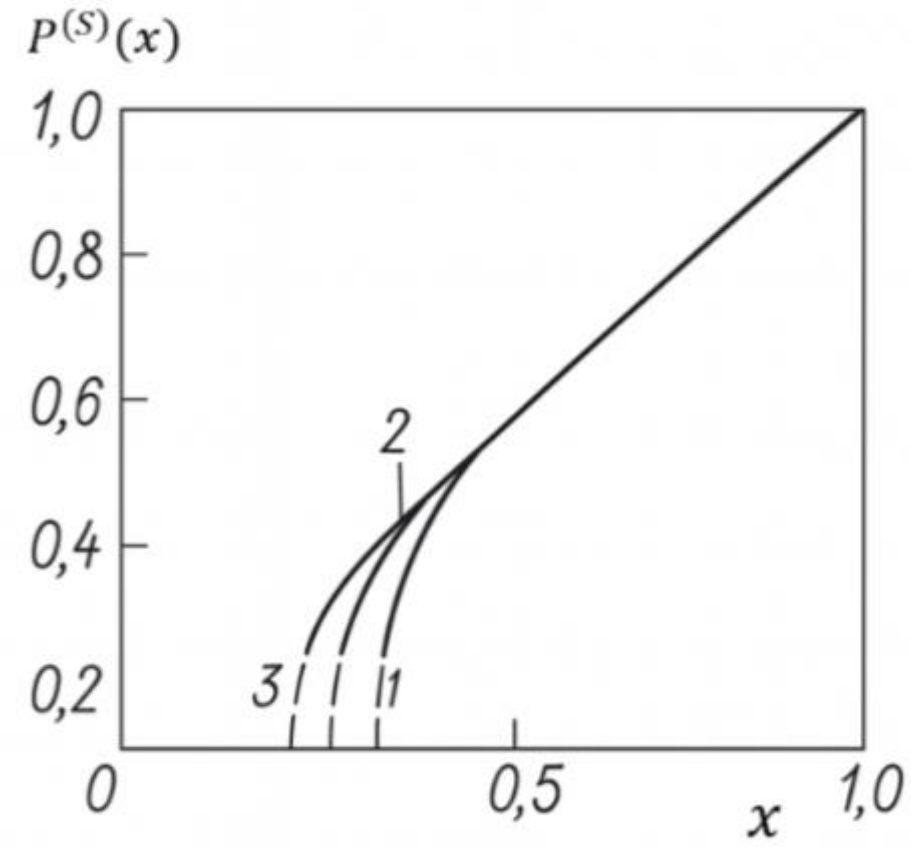

**Fig. 5.** Density of the infinite cluster $P^{(S)}(x)$ for three cubic lattices [5].

Let us discuss a specific application of heat percolation (of the phonon fluid), bearing in mind that the aim of radiation exposure in this case is to disrupt heat percolation in order to obtain an ideal thermal insulator. We will refer to Figs. 3 and 6.

Proceeding from Fig. 3, an extremely important conclusion must be drawn for our further theoretical reasoning. Indeed, considering the two dashed squares, a small one and a large one, the latter symbolizing the sample size, we can conclude that when the sample size is much larger than the correlation length L(x) for percolation, percolation occurs with high probability. When, on the contrary, the sample size is substantially smaller than the correlation length, percolation with high probability does not occur. This corresponds to the small square at the center and at the upper left corner. As we will see later, following this position, we can hope to solve our main task: to create a material with minimal thermal conductivity, that is, to prevent a through heat flux from one end of the sample to the other.

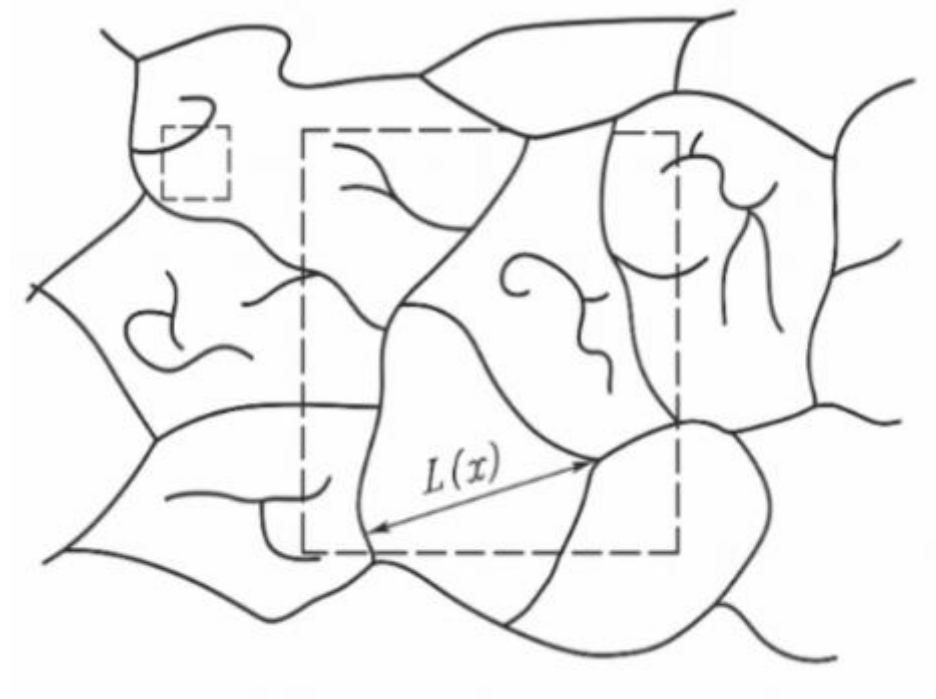

**Fig. 6.** Schematic representation of the infinite cluster and large (critical) finite clusters at $0 < x - x_c \ll 1$.

**Modeling of objects after treatment in the LSF.** Our ultimate task is to build a mechanism by which thermal conductivity through the entire sample will be low. If we turn to percolation physics (Figs. 4-6), then in the random-bond model the percolation laws are as follows: the notion of a correlation length is introduced, being the average distance between nodes of the entire chain structure. It must be compared with the size of the object in which heat percolation is expected. The following regimes are then obtained: if the object size is much larger than the correlation length, percolation proceeds successfully (see the center of Fig. 5-6, where the object size is represented by the dashed lines of a square). There is also a percolation regime in which the correlation length exceeds the object size; then percolation is suppressed (see Fig. 6, the small dashed square in the upper left corner). Thus, in this scheme, our task of suppressing the flow of the phonon fluid looks as follows. Let the initial state of the fiber network be such that its correlation length is smaller than the object size: $L_{ob} > L_{cor}$. That is, percolation takes place and thermal conductivity is high. How can one switch to a regime of low thermal conductivity? One can proceed as follows: it is necessary to randomly break the network in many places; we will then have a network with a larger cell size, which means that the average correlation length will increase: $\hat{L}_{cor} > L_{cor}$. If after the breaks the chain has $\hat{L}_{cor} > L_{ob}$ then the percolation of the phonon fluid is suppressed, that is, thermal conductivity drops sharply and almost vanishes (Fig. 5). Proceeding from this result, we must pose the question as follows: how can these breaks be made using LSF technology? We believe that the second component, powders (sintered or molten), getting into the voids of the fracton chain, can chemically interact with the long sections of fibers between nodes. If the interaction reaction is thermal, then the probability of a local break will be determined by the Boltzmann factor.

$$W_{br} \propto exp(-Q_{br}/kT^{*}) \qquad (7)$$

where $Q_{br}$ is the activation energy of fiber breaking upon interaction of the fiber with the hot powder (or its solution), $T^{*}$ is the melt temperature, and $W_{br}$ is the break probability. Obviously, the increase in $L_{cor}$ will be proportional to the indicated exponential. The quantity $Q_{br}$ can be lowered by localizing on one of the bonds an electronic excitation generated by the ionizing action of a photon from the LSF; at the same time, $T^{*}$ is the average heating temperature of the batch, or the same temperature with the addition of infrared vibrations corresponding to some bond, likewise upon absorption of an IR photon. The idea expressed here, of the simultaneous action of photochemical and thermal-chemical processes under LSF radiation, represents an important process of radiation synergism.

**Conclusion**

1. The fracton medium is an adequate model for describing the radiation response to the complex action of intense radiation.
2. Fractal aggregates formed in a fracton medium exhibit the property of shock-wave generation and nonlinear percolation.
3. The process of the initiation of thermal explosions in a fracton medium under multifactor irradiation is described using the Umov–Poynting vector as interpreted by P. L. Kapitsa.